# The Low-Temperature Nucleation Rate Anomaly in Silicate Glasses is an Artifact

Xinsheng Xia[a], D.C. Van Hoesen[b], Matthew E. McKenzie[c], Randall E. Youngman[c], K. F. Kelton[a,b,*]

[a] Institute of Materials Science and Engineering, Washington University, St. Louis, MO 63130, USA
[b] Department of Physics, Washington University, St. Louis, MO 63130, USA
[c] Science and Technology Division, Corning Incorporated, Corning, NY 14831, USA

*Corresponding author: Kenneth F. Kelton, kfk@wustl.edu

## Abstract

Over the past 40 years measurements of the nucleation rates in a large number of silicate glasses have indicated a breakdown in the widely used Classical Nucleation Theory (CNT) for temperatures below that of the peak nucleation rate. The data show that instead of steadily decreasing with decreasing temperature, the work of critical cluster formation enters a plateau, and even starts to increase. While many explanations have been offered to explain this anomaly, none have provided a satisfactory answer. We present the first experimental results that demonstrate that the anomaly is not real, but is instead an artifact arising from an insufficient annealing time at the low temperatures. The time-dependent nucleation rate was measured in a $5BaO \cdot 8SiO_2$ glass at a temperature 50 K below the peak nucleation rate temperature for a time many times longer than any previous measurement time for a silicate glass. The new data give results that are consistent with the predictions of the CNT. Since the artifact has been widely observed in many silicate glasses, these results indicate that much of the existing nucleation rate data at low temperatures are incorrect.

## Keywords

silicate glass, nucleation, critical work of cluster formation

## 1. Introduction

The development of more quantitative models for nucleation in silicate glasses is critical for accelerating the production of new glasses and glass ceramics with tailored microstructures.[1,2] Experimental studies made over the past four decades in many silicate glasses have shown that the measured time-dependent nucleation rates at temperatures below the temperature of the maximum nucleation rate contradict the predictions of the Classical Nucleation Theory (CNT).[3–7] Theoretically, the critical work of cluster formation (nucleation barrier) should decrease monotonically with decreasing temperature.[4,7] However, as shown in Fig. 1 the experimental results indicate that at low temperatures the nucleation barrier levels off or even increases with decreasing temperature.[6–15] There have been several attempts to explain this low-temperature anomaly.[6–8,16] Abyzov *et al.* showed that the anomaly cannot be explained as an elastic strain energy effect.[6] Fokin *et al.* argued, that it could be explained by adjusting the volume of the structural unit at different nucleation temperatures.[7] Gupta *et al.*[16] suggested that the size of the cooperatively rearranging regions could be the reason for the low-temperature nucleation anomaly. Abyzov *et al.*[8] proposed spatial heterogeneities, where nucleation proceeds only in liquid-like regions. Recently, by reanalyzing published experimental data, Cassar *et al.* suggested that the low-temperature anomaly could be an experimental artifact arising from an inadequate time at low temperature to reach steady-state.[17] However, their conclusion was based largely on a statistical analysis of the existing data. While they also presented three new measurements of nucleation in a lithium disilicate glass, the steady-state nucleation rate was not reached in their measurement below the peak nucleation temperature. There are, then, no existing experimental data that can directly confirm or refute the assertion of Cassar *et al.*, nor the proposed explanations for the anomaly.

Instead of continuing the search for possible reasons for the low-temperature nucleation anomaly, we therefore took a step back to experimentally check if the low-temperature nucleation anomaly widely found in the earlier sudies[6–15] was true or not. The time-dependent nucleation rate was measured in $5BaO{\cdot}8SiO_2$ glasses that were held at a nucleation temperature of 948 K, which is 50 K below the temperature of the maximum nucleation rate, for up to 115 days. This time is much longer than any used in the earlier studies of different silicate glasses.[9–15,18–20] A barium-silicate glass was chosen since they have larger nucleation rates than other glasses, such as lithium disilicate or soda-lime silicate, so that it takes less time to obtain a significant number of nuclei. The crystals in the $5BaO{\cdot}8SiO_2$ glasses are spherical, making it easier to accurately measure the nuclei density than in the $BaO{\cdot}2SiO_2$ glass, for example, where the crystals have irregular shapes.[9] The experimental data obtained confirm the suggestion by Cassar *et al.*[17] that the nucleation anomaly at low temperatures is not real, but is rather an experimental artifact due to the short nucleation times used in earlier studies. Instead of plateauing or increasing with decreasing temperatures for temperatures below the peak nucleation temperature, the critical work of cluster formation monotonically decreases with decreasing temperature, following the trend expected from the Classical Nucleation Theory.

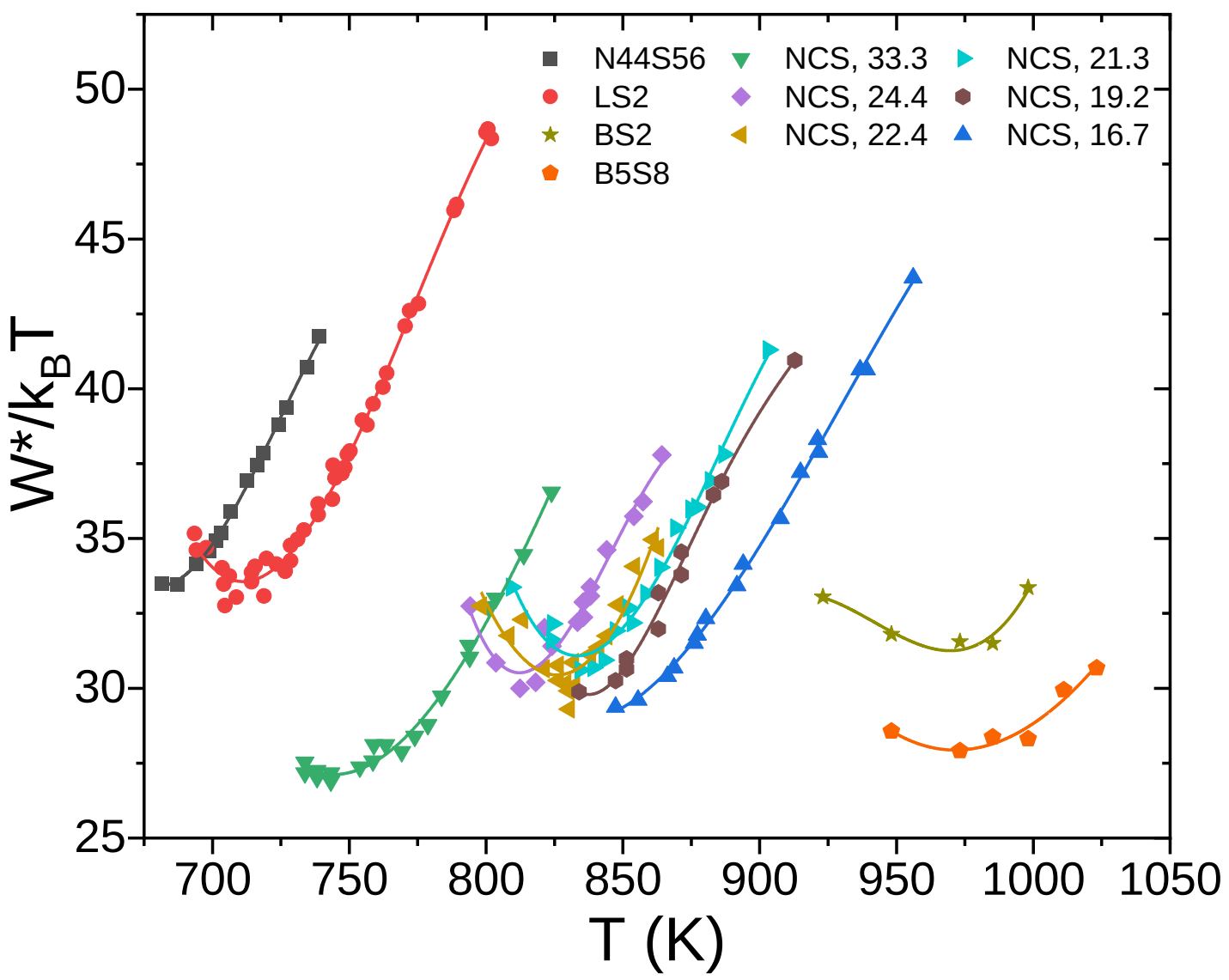

**Figure 1** The nucleation barrier, $W^*$ (scaled to $k_BT$, where $k_B$ is Boltzmann's constant) as a function of temperature ($T$) for silicate glasses. These data are for the following glasses: $44Na_2O{\cdot}56SiO_2$ (N44S56)[7,15], $Li_2O{\cdot}2SiO_2$ (LS2)[7,10], $BaO{\cdot}2SiO_2$ (BS2)[9], $5BaO{\cdot}8SiO_2$ (B5S8)[9], and $xNa_2O{\cdot}(50\text{-}x)CaO{\cdot}50SiO_2$ (NCS) where x values are 33.3[7,14], 24.4[6,12], 22.4[6,12], 21.3[6,12], 19.2[6,12], and 16.7[7,11–13].

## 2. Results

The approach used to measure the nucleation rate is discussed in the Method section; the results are discussed here. Figure 2 shows the measured number of nuclei per unit volume, $N_V$, as a function of nucleation time at 948 K, together with data measured for this same glass earlier.[9] Initially $N_V$ increases nonlinearly with time; it eventually becomes linear with time, indicating that steady-state has been reached. The steady-state nucleation rate ($I^{st}$) and the induction time ($\theta_{n^*(T_G)}$) are obtained from the slope and intercept with the time axis, respectively, of the linear portion of the curve.[21] The measured values are $I^{st}$ (= 397 ± 18 $mm^{-3}s^{-1}$) and $\theta_{n^*(T_G)}$ (= 39,638.5 ± 3229.3 min.). These values are listed in Table 1, together with our previous results.[9] With the significantly longer nucleation time, the new values of $I^{st}$ and $\theta_{n^*(T_G)}$ are 7 times and 111 times, respectively, larger than the values obtained in the previous study.[9]

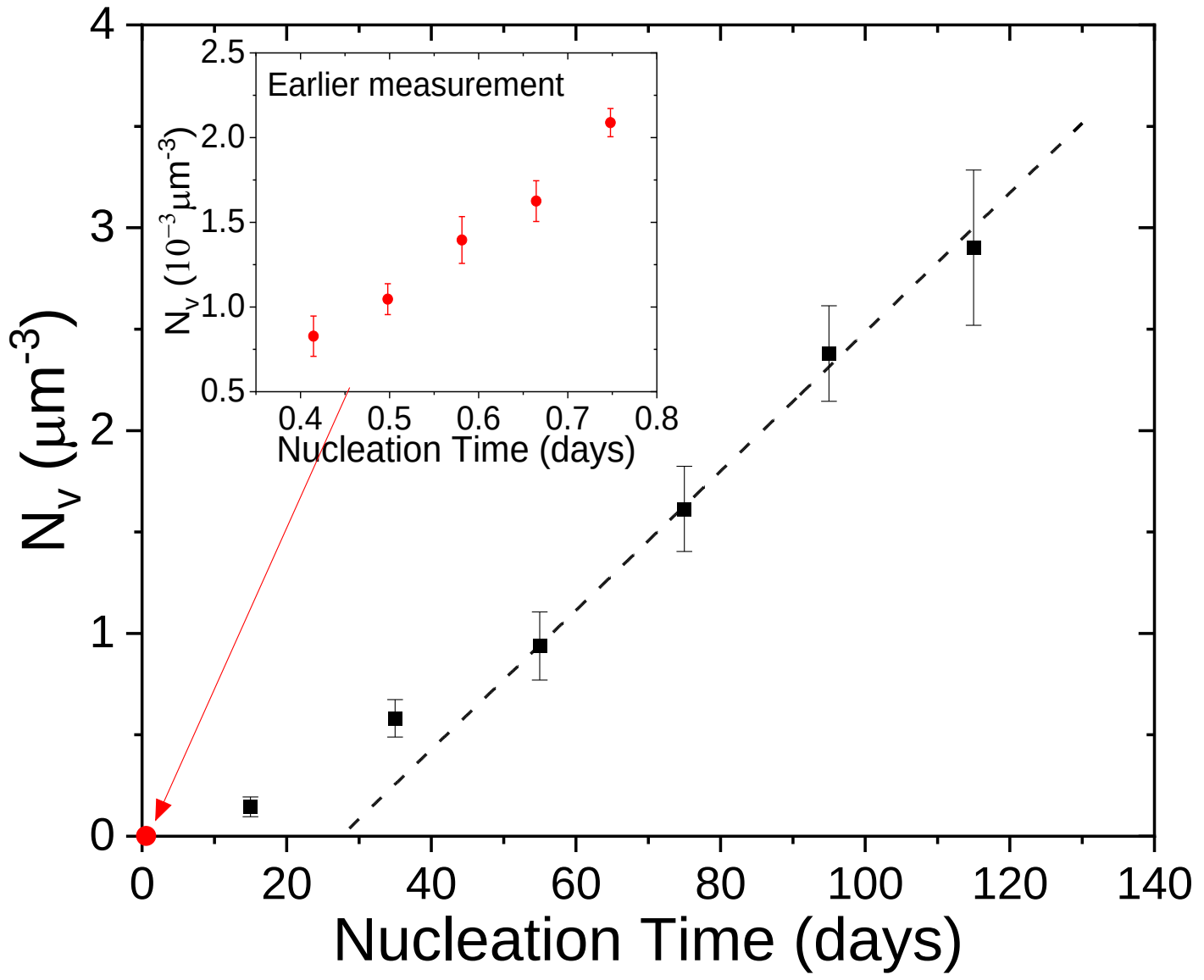

**Figure 2** $N_V$ as a function of nucleation time at 948 K for the $5BaO{\cdot}8SiO_2$ glasses. The black points are the new data obtained here and the red points are the data from an earlier study[9] (see inset). The dashed lines show the linear fits in the steady-state range. (The error bars indicate the standard deviation of $N_V$.)

**Table 1**

Steady-state rates and induction times for nucleation in $5BaO{\cdot}8SiO_2$ glasses

| **Temperature, *T* (K)** | **Steady State Nucleation Rate, $I^{st}$ (mm$^{-3}$s$^{-1}$)** | **Induction Time, $\theta_{n^*(T_G)}$ (minutes)** |
|---|---|---|
| 948 (this measurement) | 397 ± 18 | 39638.5 ± 3229.3 |
| 948* | 48 ± 3 | 354 ± 41 |
| 973* | 746 ± 72 | 45 ± 4 |
| 985* | 1345 ± 25 | 16.1 ± 0.4 |
| 998* | 3135 ± 54 | 7.4 ± 0.3 |
| 1011* | 2599 ± 127 | 1.8 ± 0.2 |
| 1023* | 2035 ± 28 | 1.1 ± 0.1 |
| 1048* | 669 ± 53 | Not Determined |

Note: 948 K (this measurement) is the measurement here using 1073 K as the growth temperature. All the data labeled with * are from our previous study[9], which used 1119 K as the growth temperature. The value and standard error were determined from the linear fit in the $N_V$ vs. nucleation time plots using the instrumental weighting in Origin software.

The methods used to obtain the interfacial free energy $\sigma$, the critical work of cluster formation $W^*$, and the induction time for the critical size at the nucleation temperature $\theta_{n^*(T_N)}$ from the nucleation data are discussed in the Supplementary Methods section in the Supplementary Information file. The values for $I^{st}$ and $\theta_{n^*(T_G)}$ at 948 K from this study were combined with values obtained at temperatures at or above the temperature for the maximum steady-state nucleation rate, whose $I^{st}$ and $\theta_{n^*(T_G)}$ are known and reported by Xia et al.[9] The measured induction time corresponds to that for the critical size at the growth temperature, $\theta_{n^*(T_G)}$. To compare with predictions of CNT, the critical size at the nucleation temperature, $\theta_{n^*(T_N)}$, is required. This was computed from $\theta_{n^*(T_G)}$ following a method discussed earlier. [9] The Turnbull approximation[9,22] was used to calculate the driving free energy as a function of temperature, $|\Delta g_V|$, assuming one unit of $5BaO{\cdot}8SiO_2$ (fig. 3.a). The calculated interfacial free energy, $\sigma$, is shown in fig. 3.b( the details of how $\sigma$ was calculated are given in the Supplementary Methods section in the Supplementary Information file), along with the values obtained previously[9]. The previous results showed that while at high temperature $\sigma$ decreases linearly with decreasing temperature, this changed to an increasing $\sigma$ with decreasing temperature for temperatures below the temperature for maximum nucleation rate (998 K). The new measurements obtained here show that $\sigma$ monotonically decreases with decreasing temperature over the whole temperature range, consistent with the predictions of the Diffuse Interface Theory of nucleation.[23–26] Also unlike the previous results[9] (fig. 3.c), $W^*/k_BT$ decreases over the entire temperature range, rather than decreasing with decreasing temperature only when the temperature is higher than the peak nucleation temperature (998 K), but plateauing at lower temperatures. The new results follow the trend expected from CNT. Finally, CNT predicts that a plot of $\ln\left(I^{st}\theta_{n^*(T_N)}\right)$ as a function of $1/\left(T\left|\Delta g_v\right|^2\right)$ should be linear[3] when $\sigma$ is a constant or the relative change in $\sigma$ as a function of temperature is smaller than the relative change in $\left|\Delta g_V\right|$ as a function of temperature. As shown in fig 3.d, this is true if the new data are used, as opposed with the previous results showing a significant departure from the straight line behavior.[9]

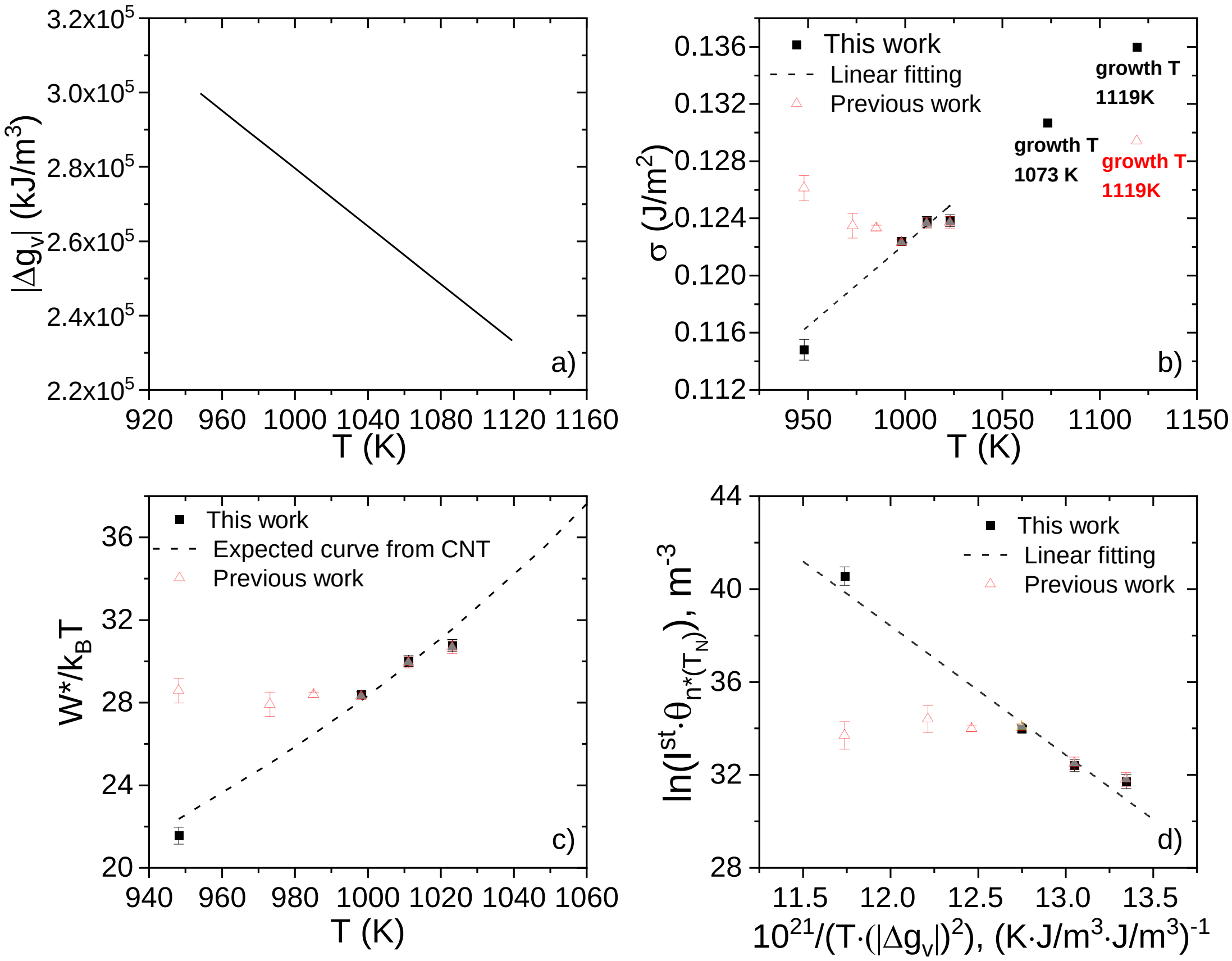

**Figure 3** – The values of $\sigma$, $W^*/k_BT$ and $\ln\left(I^{st}\theta_{n^*(T_N)}\right)$ from this study, using the longer nucleation time at 948 K, and from the previous study[9] using a shorter-time nucleation data at low temperatures for $5BaO{\cdot}8SiO_2$ glasses. (a) The calculated $\left|\Delta g_V\right|$ as a function of temperature; (b) the calculated interfacial free energy, $\sigma$, as a function of temperature; (c) $W^*/k_BT$ as a function of temperature; (d) the natural logarithm of the product of the steady-state nucleation rate and the induction time for the critical size at the nucleation temperature, $\ln\left(I^{st}\theta_{n^*(T_N)}\right)$ as a function of $1/\left(T\left|\Delta g_v\right|^2\right)$. The errors were calculated using the 95% confidence intervals of $I^{st}$ and $\theta_{n^*(T_G)}$. The red symbols represent the values obtained in the previous study.[9]

The nucleation rate as a function of temperature was calculated assuming CNT and using the values for $|\Delta g_V|$ shown in fig. 3.a and σ given by the dashed line in fig. 3.b, and assuming the Kashchiev expression[27] to calculate the diffusion coefficient from the induction time for the critical size at the nucleation temperature. The result is shown by the solid line in fig. 4; the 95% confidence bounds are indicated by the dashed lines. Except for the data point at the highest temperature (shown in red), the high temperature data and the new low temperature data point (shown in black) agree reasonably well with the calculated nucleation rates. Importantly, the two data points at 973 K and 985 K fall below or close to the lower limit of the 95% confidence bounds, indicating that they have not yet achieved the steady-state value. The data point at 1048 K falls outside of the higher limit of the confidence bounds. This likely is an artifact of the fit, however. The induction time was not measurable at this temperature; instead, it was estimated from the data at 998 K, 1011 K and 1023 K.

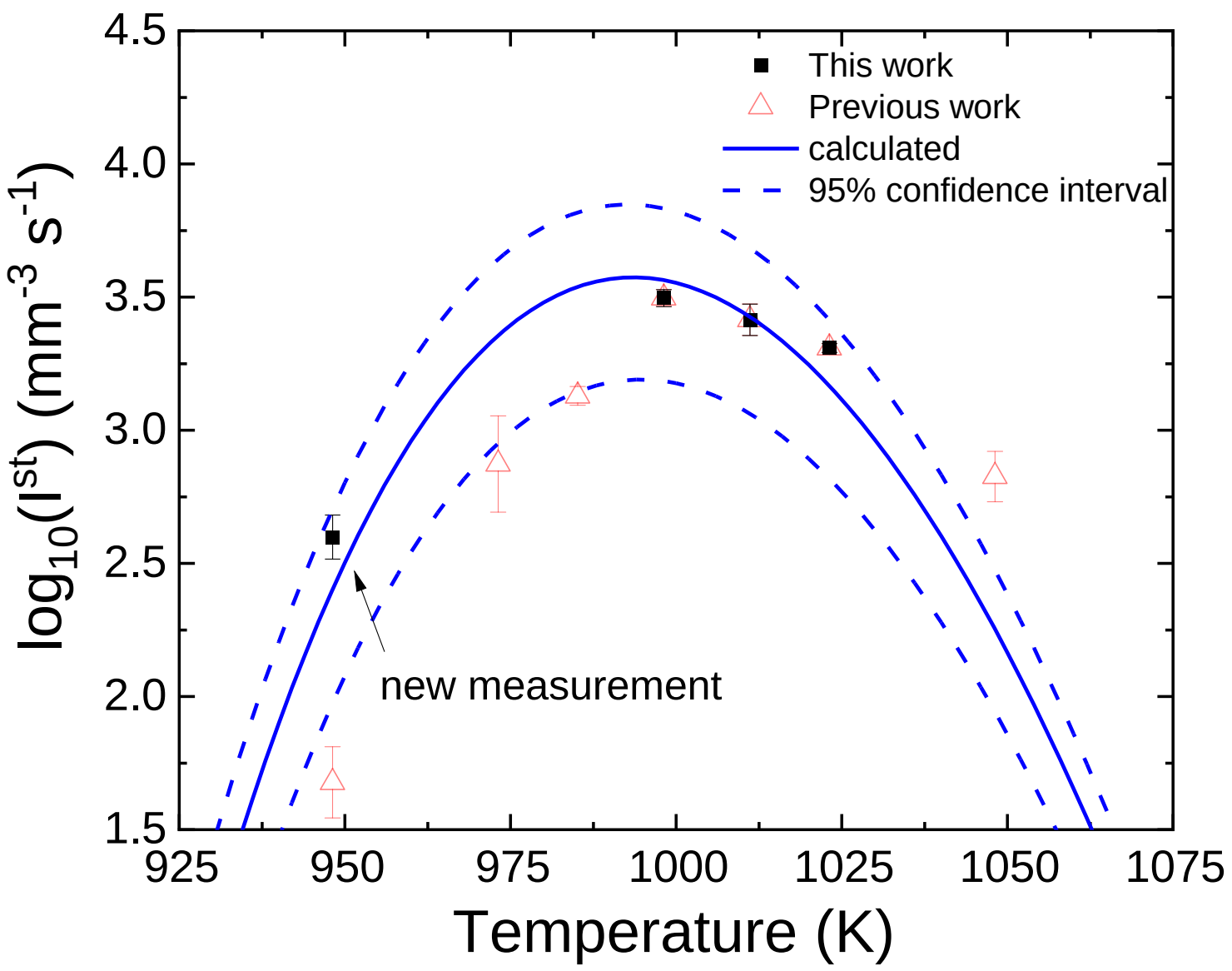

**Figure 4 –** A comparison between the calculated steady-state nucleation rate from the Classical Nucleation Theory (solid blue curve) and the measured data. The red symbols represent the values obtained in the previous study.[9] The dashed lines are the 95% confidence limits for the calculated curve. The error bars are the 95% confidence intervals of the measured data.

## 3. Conclusions

In summary, the time-dependent nucleation rate was measured in $5BaO{\cdot}8SiO_2$ glasses at a temperature that was 50 K below the peak nucleation rate temperature. Earlier measurements of the nucleation rate in this glass[9] showed an anomalous behavior at these low temperatures that was consistent with what has been reported in many other silicate glasses.[3–7] For the new measurements reported here the glasses were given a much longer nucleation treatment than was used in all previous measurements.[9–15,18–20] These new data do not show a low temperature anomaly. In contradiction to previous results, the interfacial free energy decreases with decreasing temperature over the whole measurement temperature range, consistent with predictions of the diffuse interface theory of nucleation[23–25]. Also, following the trend predicted by the Classical Nucleation Theory, the critical work of cluster formation monotonically decreases with decreasing temperature instead of plateauing or increasing with decreasing temperature for temperatures below the peak nucleation temperature, which the earlier studies showed. These results demonstrate that the anomaly is not real, but is an experimental artifact due to insufficient nucleation treatment times at low temperatures in the previous studies. Based on this result and given the practical importance of knowing the nucleation rate as a function of temperature, the low temperature data in silicate glasses should be re-measured since they are likely incorrect.

## Methods

The $5BaO{\cdot}8SiO_2$ glasses were prepared by Corning Incorporated using the melting and quenching procedures discussed by Xia *et al.*[9]. The source materials were barium carbonate and silica. The composition of the prepared bulk glasses were measured by Inductively Coupled Plasma – Optical Emission Spectroscopy (ICP-OES) to be BaO (38.73 mol%), $SiO_2$ (61.21 mol%), SrO (0.04 mol%), $Fe_2O_3$ (0.01 mol%), and $Al_2O_3$ (<0.01 mol%). As reported earlier[9] the measured glass transition temperature for the $5BaO{\cdot}8SiO_2$ glass is 970 K. Prior to the heat treatments, the bulk $5BaO{\cdot}8SiO_2$ glasses were cut into plates having an area of approximately 3.8 mm x 3.0 mm and a thickness of 0.98 ± 0.07 mm.

The time-dependent nucleation rate was measured using the two-step heating method[19,28]. Samples were first heated at a temperature where the nucleation rate is large, but the growth velocity is small. These nuclei were then grown to observable size by heating at a temperature where the growth velocity is larger than that at nucleation temperature but the nucleation rate is small. During the nucleation treatment the samples were heated together in a container (a 5mL Coors high alumina combustion boat, Sigma Aldrich) in a Lindberg tube furnace at 948 ± 2 K (the temperature range of the center of the furnace). To mitigate possible diffusion between the samples and the container, an additional spacer of $5BaO{\cdot}8SiO_2$ glass (~ 1cm thick) was placed between the two. The spacer was replaced with a new one every 25 or 30 days. When each target heating time was reached, the collection of samples and container were taken out of the furnace, air quenched onto a metal plate to room temperature, and one sample was randomly removed. The remaining samples were then reinserted into the furnace and positioned close to the center of the 948 ± 2 K temperature range in the furnace. Samples were nucleated for 15, 35, 55, 75, 95 and 115 days. The nuclei density in these samples, which had been held at the nucleation temperature for a much longer time than in previous studies,[9–15,18–20] was so large that due to crystal impingement they could not be grown to sizes that could be observed in optical microscopy. Instead, a growth treatment was selected that produced crystals with diameters smaller than one micrometer; the nuclei density was then measured in a Scanning Electron Microscope (SEM). After the nucleation treatment, the samples were placed inside a 5mL Coors high alumina combustion boat (Sigma Aldrich) and inserted together into a Lindberg Blue M three-zone tube furnace that had been equilibrated at 1073 K. Eight minutes after insertion, the samples and the boat were removed from the furnace and air quenched onto a metal plate. The number of the new nuclei formed during the growth treatment was negligible compared with the number of nuclei created during the nucleation treatment.

After the nucleation and growth heat treatments the samples were polished, etched, and cleaned following the same procedures used previously.[9] At least 250 μm thickness of the sample surfaces were removed during polishing. The nuclei density was determined by imaging the sample surface using a a Thermofisher Quattro S Environmental SEM with a 10 kV accelerating voltage, 30 Pa chamber pressure, and a low-vacuum detector operating in the secondary electron mode. Multiple

SEM images were taken from each sample. Typical SEM images showing spherical crystals are shown in Supplementary Figure 1 in the supplementary information file. For each image the number of crystals per area, $N_S$, and the average of the reciprocal diameters, $\bar{Y}$, were measured. The number of crystals per unit volume, $N_V$, was determined using[29,30]

$$N_V = \frac{2}{\pi} N_S \bar{Y} \,. \tag{1}$$

For each sample, the standard deviation for $N_V$ was calculated from the multiple images. The microscopy resolution limit-related correction for a monodispersed system[31] and the density of nuclei in the as-quenched glass were used to further correct $N_V$.

**Acknowledgements**

The authors thank E. Zanotto and D. Cassar for pointing out their statistical analysis and for very useful discussions. We also thank M. Sellers, R. Chang, A. Gangopadhyay, and P. Gibbons for useful discussions. We acknowledge the Institute of Materials Science and Engineering at Washington University in St. Louis for the use of the Thermofisher Quattro S Environmental SEM. This work is financially supported by NSF GOALI grant DMR 17-20296, and by Corning Incorporated.

**Author Contributions**

K.F.K. conceived of the idea for this study. X.X. detailed the experiment plan, conducted the nucleation measurements and analyzed the data. D.C.V. assisted with the calibration of furnace, the etching process, and the data analysis. All authors engaged in useful discussions during the ongoing study and contributed to the writing and editing of the paper.

**Competing Interests Statement**

The authors declare no competing financial interests.

## References

1. Mauro, J. C. Grand challenges in glass science. *Front. Mater.* **1**, 1–5 (2014).

2. Mauro, J. C., Philip, C. S., Vaughn, D. J. & Pambianchi, M. S. Glass science in the United States: Current status and future directions. *Int. J. Appl. Glas. Sci.* **5**, 2–15 (2014).

3. Weinberg, M. C. & Zanotto, E. D. Re-examination of the temperature dependence of the classical nucleation rate: Homogeneous crystal nucleation in glass. *J. Non. Cryst. Solids* **108**, 99–108 (1989).

4. Schmelzer, J. W. P., Fokin, V. M. & Abyzov, A. S. Crystallization of Glass: What We Know, What We Need to Know. *Int. J. Appl. Glas. Sci.* **7**, 253–261 (2016).

5. Fokin, V. M., Zanotto, E. D., Yuritsyn, N. S. & Schmelzer, J. W. P. Homogeneous crystal nucleation in silicate glasses: A 40 years perspective. *J. Non. Cryst. Solids* **352**, 2681–2714 (2006).

6. Abyzov, A. S., Fokin, V. M., Rodrigues, A. M., Zanotto, E. D. & Schmelzer, J. W. P. The effect of elastic stresses on the thermodynamic barrier for crystal nucleation. *J. Non. Cryst. Solids* **432**, 325–333 (2016).

7. Fokin, V. M. *et al.* Crystal nucleation in glass-forming liquids: Variation of the size of the 'structural units' with temperature. *J. Non. Cryst. Solids* **447**, 35–44 (2016).

8. Abyzov, A. S., Fokin, V. M., Yuritsyn, N. S., Rodrigues, A. M. & Schmelzer, J. W. P. The effect of heterogeneous structure of glass-forming liquids on crystal nucleation. *J. Non. Cryst. Solids* **462**, 32–40 (2017).

9. Xia, X. *et al.* Time-dependent nucleation rate measurements in $BaO\cdot2SiO_2$ and $5BaO\cdot8SiO_2$ glasses. *J. Non. Cryst. Solids* **525**, 119575 (2019).

10. Nascimento, M. L. F., Fokin, V. M., Zanotto, E. D. & Abyzov, A. S. Dynamic processes in a silicate liquid from above melting to below the glass transition. *J. Chem. Phys.* **135**, (2011).

11. Potapov, O. V., Fokin, V. M., Ugolkov, V. L., Suslova, L. Y. & Filipovich, V. N. Influence of $Na_2O$ content on the nucleation kinetics in glasses of compositions close to the $Na_2O\cdot2CaO\cdot3SiO_2$ stoichiometry. *Glas. Phys. Chem.* **26**, 27–32 (2000).

12. Yuritsyn, N. S. Nucleation of crystals in sodium-calcium-silicate glasses of the metasilicate section. *Glas. Phys. Chem.* **41**, 112–115 (2015).

13. Gonzalez-Oliver, C. J. R. & James, P. F. Crystal nucleation and growth in a $Na_2O\cdot2CaO\cdot3SiO_2$ glass. *J. Non. Cryst. Solids* **38–39**, 699–704 (1980).

14. Kalinina, A. M., Filipovich, V. N. & Fokin, V. M. Stationary and non-stationary crystal nucleation rate in a glass of $2Na_2O\cdot CaO\cdot3SiO_2$ stoichiometric composition. *J. Non. Cryst. Solids* **38&39**, 723–728 (1980).

15. Fokin, V. M. & Yuritsyn, N. S. The nucleation and growth rates of sodium metasilicate crystals in sodium silicate glass $44Na_2O\cdot56SiO_2$. *Glas. Phys. Chem.* **23**, 236–239 (1997).

16. Gupta, P. K., Cassar, D. R. & Zanotto, E. D. Role of dynamic heterogeneities in crystal nucleation kinetics in an oxide supercooled liquid. *J. Chem. Phys.* **145**, (2016).

17. Cassar, D. R., Serra, A. H., Peitl, O., Rodrigues, A. M. & Zanotto, E. D. The failure of the Classical Nucleation Theory at low temperatures resolved. *arXiv Prepr.* **arXiv:1902**.03193 (2019).

18. Zanotto, E. D. & James, P. F. Experimental tests of the classical nucleation theory for glasses. *J.*

*Non. Cryst. Solids* **74**, 373–394 (1985).

19. Narayan, K. L. & Kelton, K. F. First measurements of time-dependent nucleation as a function of composition in $Na_2O{\cdot}2CaO{\cdot}3SiO_2$ glasses. *J. Non. Cryst. Solids* **220**, 222–230 (1997).

20. Deubener, J., Brückner, R. & Sternitzke, M. Induction time analysis of nucleation and crystal growth in di- and metasilicate glasses. *J. Non. Cryst. Solids* **163**, 1–12 (1993).

21. Kelton, K. F. & Greer, A. L. *Nucleation in condensed matter: applications in materials and biology*, Elsevier, Amsterdam (2010).

22. Turnbull, D. Formation of crystal nuclei in liquid metals. *J. Appl. Phys.* **21**, 1022–1028 (1950).

23. Spaepen, F. Homogeneous Nucleation and the Temperature Dependence of the Crystal-Melt Interfacial Tension. *Solid State Phys. - Adv. Res. Appl.* **47**, (1994).

24. Gránásy, L. Diffuse interface theory of nucleation. *J. Non. Cryst. Solids* **162**, 301–303 (1993)..

25. Gránásy, L. Diffuse interface model of crystal nucleation. *J. Non. Cryst. Solids* **219**, 49–56 (1997).

26. Turnbull, D. Kinetics of solidification of supercooled liquid mercury droplets. *J. Chem. Phys.* **20**, 411–424 (1952).

27. Kashchiev, D. Solution of the non-steady state problem in nucleation kinetics. *Surf. Sci.* **14**, 209–220 (1969).

28. James, P. F. Kinetics of crystal nucleation in lithium silicate glasses. *Phys. Chem. Glas.* **15**, 95–105 (1974).

29. DeHoff, R. & Rhines, F. Determination of number of particles per unit volume from measurements made on random plane sections: the general cylinder and the ellipsoid. *Trans. Met. Soc. AIME* **221**, 975–982 (1961).

30. Zanotto, E. D. The effects of amorphous phase separation on crystal nucleation and growth in baria-silica and lithia-silica glasses. Ph.D. thesis, University of Sheffield, (1982)

31. Zanotto, E. D. & James, P. F. A theoretical and experimental assessment of systematic errors in nucleation experiments. *J. Non. Cryst. Solids* **124**, 86–90 (1990).

## Supplementary Information

(a)

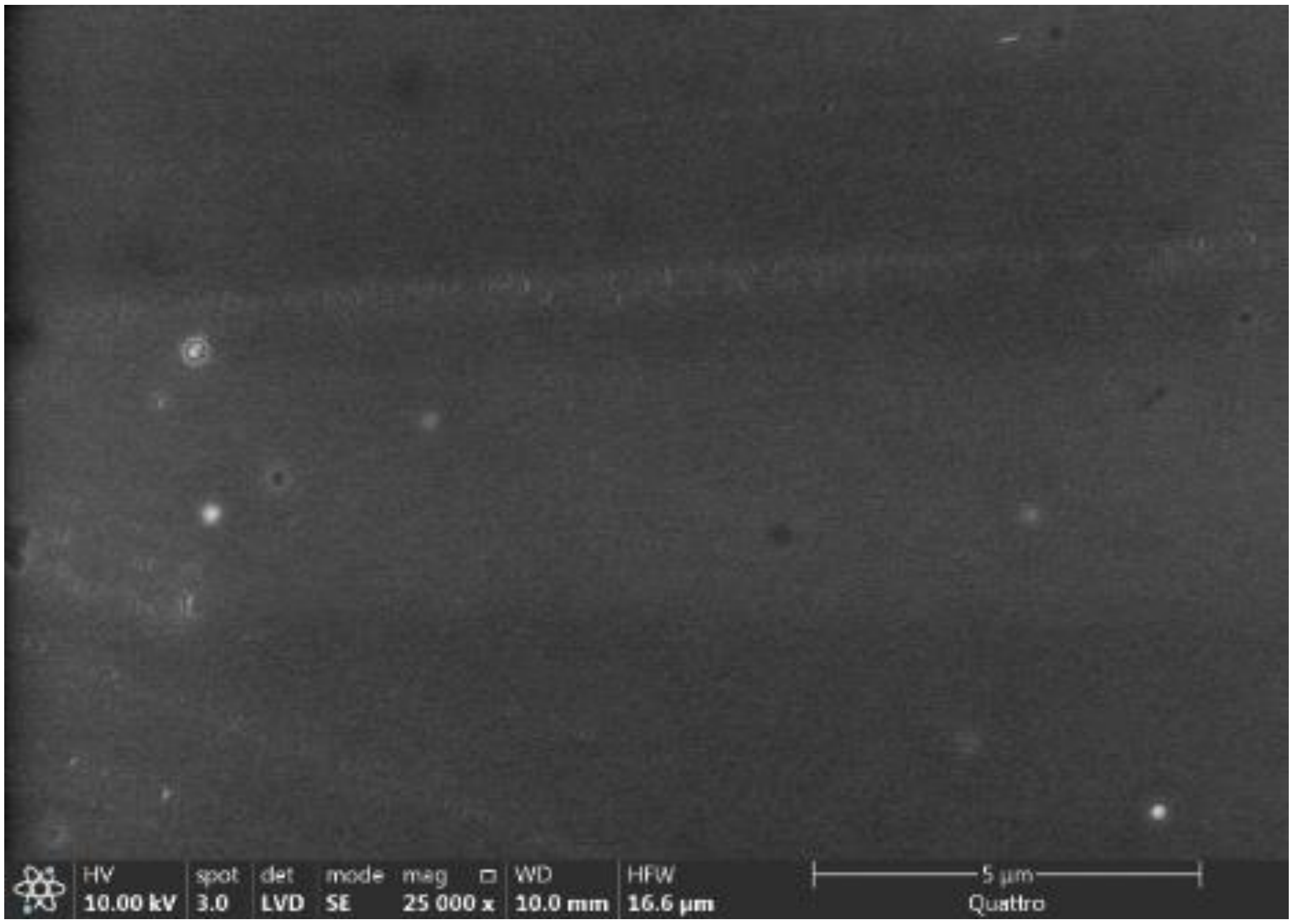

(b)

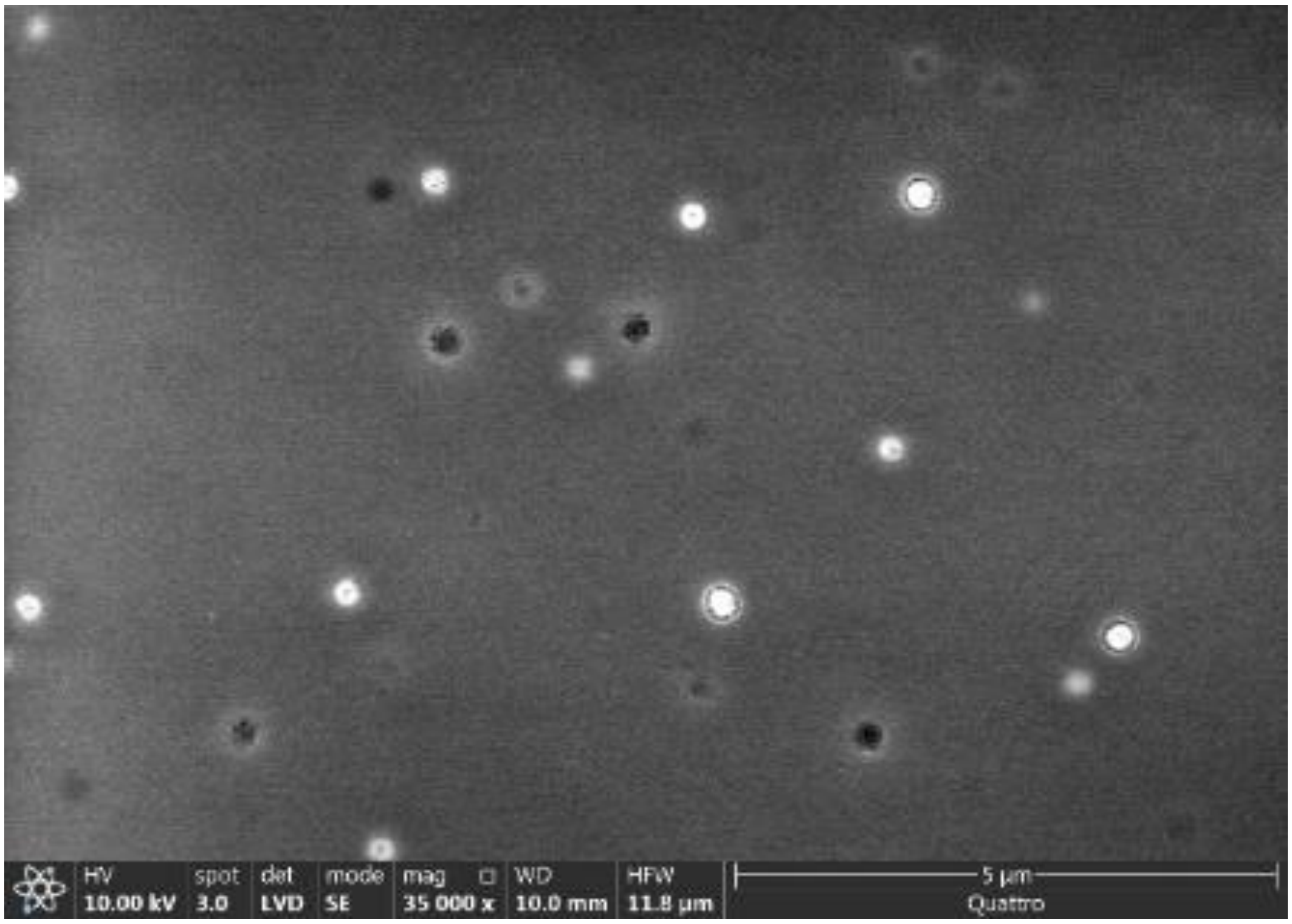

(c)

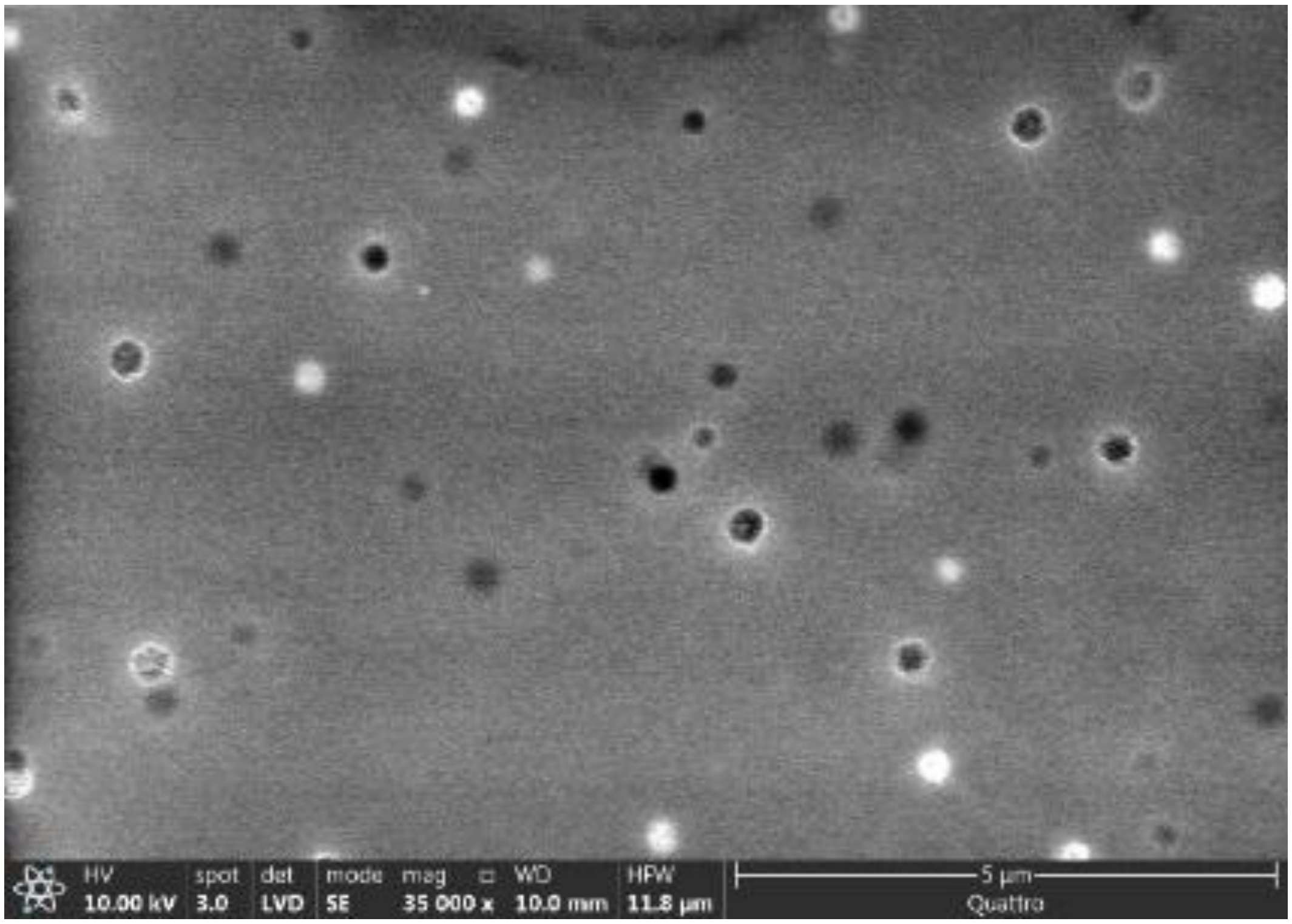

(d)

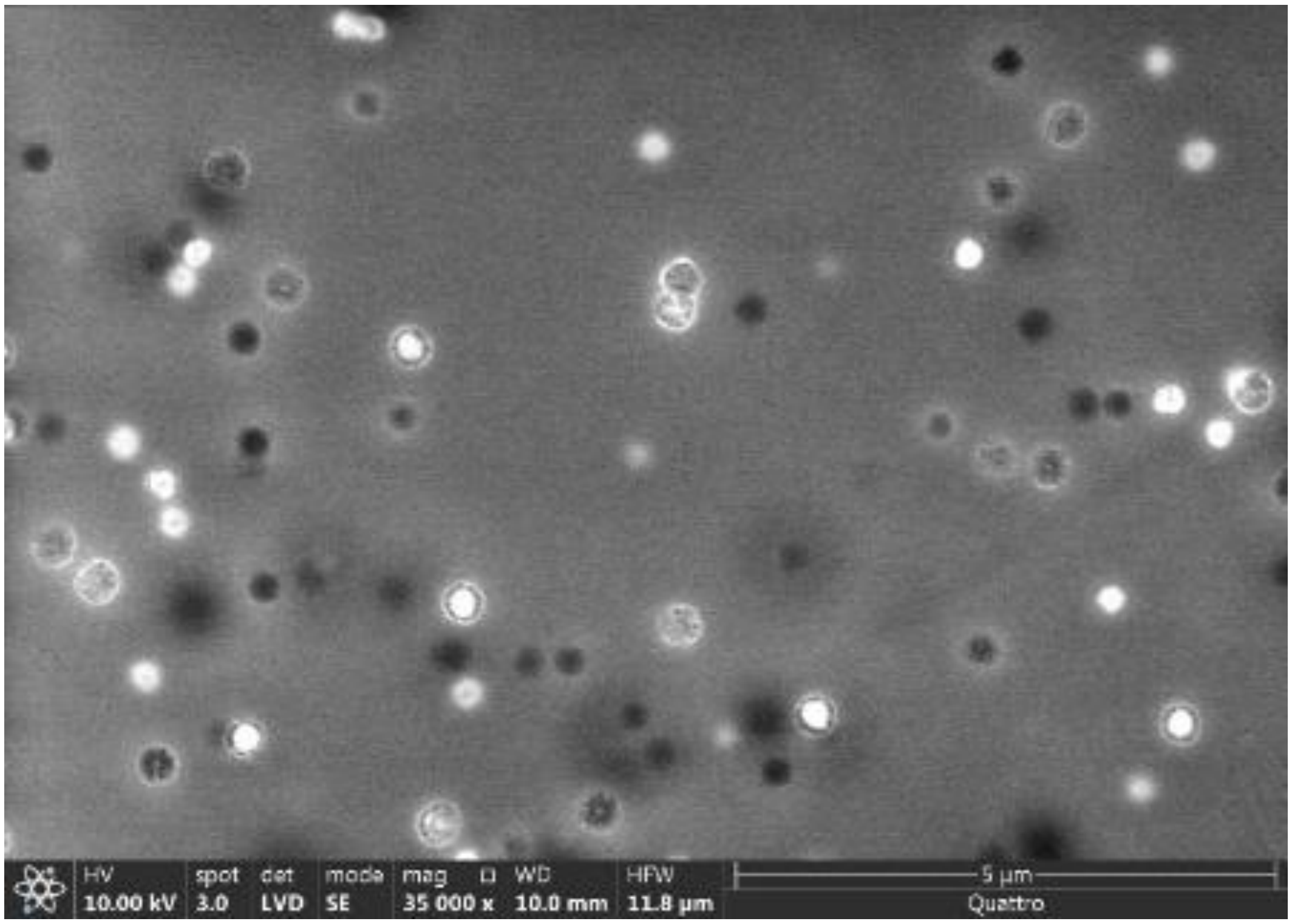

(e)

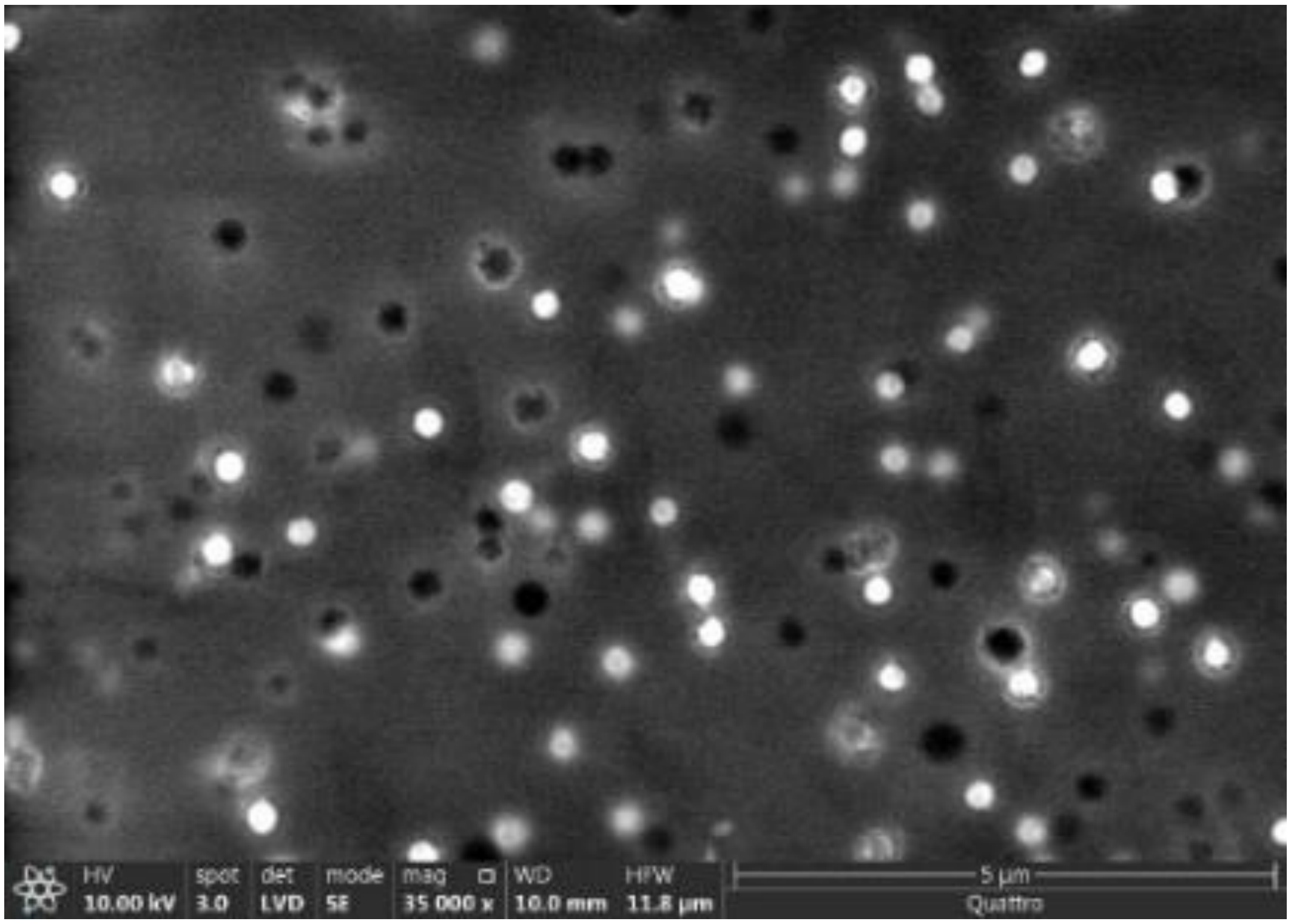

(f)

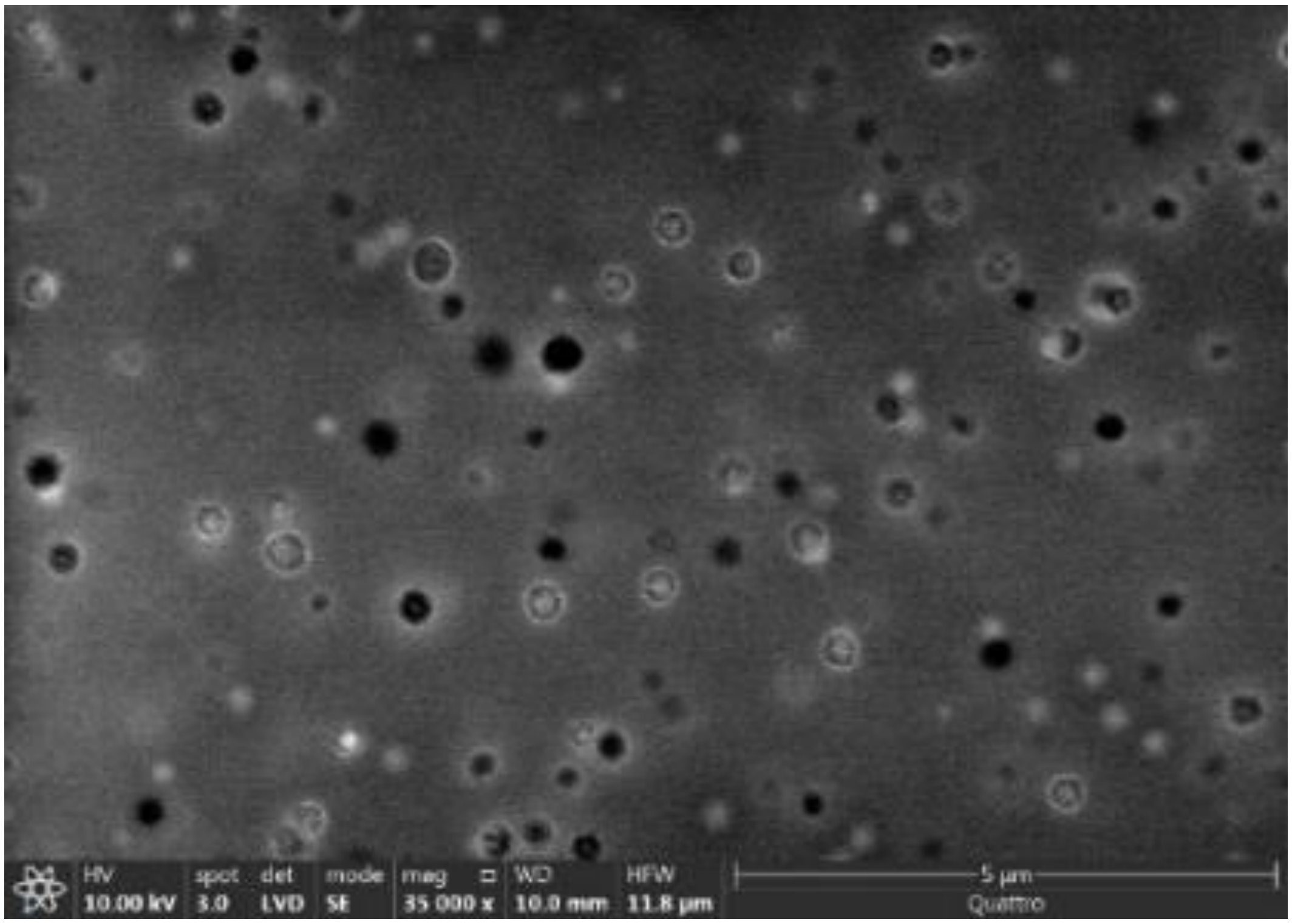

**Supplementary Figure 1** Typical SEM images obtained for $5BaO{\cdot}8SiO_2$ glasses nucleated at 948 K for (a) 15 days, (b) 35 days, (c) 55 days, (d) 75 days, (e) 95 days, (f) 115 days, and then growth treated at 1073 K.

**Supplementary Methods: the analysis methods for the interfacial free energy $\sigma$, critical work of cluster formation $W^*$, and corrected induction time $\theta_{n*(T_N)}$**

The data analysis used eq. (S-1), which was derived from the Classical Nucleation Theory (CNT) (see Xia *et al.*[1]) .

$$\ln\left(I^{st}\theta_{n^*(T_G)}\right) = \ln\left\{\left(\frac{\sigma_{T_G}}{\sigma_{T_N}}\right)\Big/\left(\frac{|\Delta g_{\mathrm{v}}|_{T_G}}{|\Delta g_{\mathrm{v}}|_{T_N}}\right) + \ln\left[\left(\frac{\sigma_{T_G}}{\sigma_{T_N}}\right)\Big/\left(\frac{|\Delta g_{\mathrm{v}}|_{T_G}}{|\Delta g_{\mathrm{v}}|_{T_N}}\right) - 1\right] + \ln\left[\frac{32\pi}{k_B}\right] + \ln\left[\frac{\left(\sigma_{T_N}\right)^3}{T_N\left(|\Delta g_{\mathrm{v}}|_{T_N}\right)^2}\right] + \zeta_{\mathrm{E}} - 2\right\}$$
$$+ \ln\left[\frac{32}{\pi^2\bar{\mathrm{v}}^2}\right] + \frac{1}{2}\ln\left(\frac{\left(\sigma_{T_N}\right)^3 k_B T_N}{\left(|\Delta g_{\mathrm{v}}|_{T_N}\right)^4}\right) - \left(\frac{16\pi}{3k_B}\frac{1}{T_N\left(|\Delta g_{\mathrm{v}}|_{T_N}\right)^2}\right)\left(\sigma_{T_N}\right)^3 \quad .$$

(S-1)

Here, $I^{st}$ is the measured steady-state nucleation rate, $\theta_{n*(T_G)}$ is the measured induction time, $\sigma_{T_N}$ is the interfacial free energy at the nucleation temperature, $\sigma_{T_G}$ is the interfacial free energy at the growth temperature, $|\Delta g_{\mathrm{v}}|_{T_N}$ and $|\Delta g_{\mathrm{v}}|_{T_G}$ are the Gibbs driving free energies per unit volume at the nucleation and growth temperature, respectively, $k_B$ is Boltzmann's constant, $T_N$ is the nucleation temperature, $\zeta_{\mathrm{E}}$ is Euler's constant (0.5772…), and $\bar{\mathrm{v}}$ is the monomer volume.[1] Since $\sigma_{T_N}$ and $\sigma_{T_G}$ must be known, the new data presented here at 948 K were analyzed using our previous measurements of $I^{st}$ and $\theta_{n*(T_G)}$ data for temperatures at and above that of the maximum steady-state nucleation rate.[1] Those high temperature data are free from the problems associated with the low temperature data.

The analysis steps were:

(a) Linearly extrapolate the interfacial free energies at 998 K, 1011 K, and 1023 K calculated in Xia *et al.*[1] to obtain an initial estimate for the interfacial free energy at the growth temperature, $\sigma_{T_G}$ (1073 K for this study).

(b) Use this value of $\sigma_{T_G}$ and eq. (S-1) to calculate the initial value of interfacial free energy at the nucleation temperature, $\sigma_{T_N}$ (948 K in this study)

(c) Linearly extrapolate the initial value of $\sigma_{T_N}$ at 948 K (this study) and $\sigma_{T_N}$ at 998 K, 1011 K, and 1023 K (Xia *et al.*[1]) to a new estimate of $\sigma_{T_G}$ at 1073 K and 1119 K(the growth temperature used in the earlier study, Xia *et al.*[1])

(d) If the difference between the new value of $\sigma_{T_G}$ and the previous value of $\sigma_{T_G}$ at either 1073 K and 1119 K is larger than $10^{-5}$ J/m$^2$, use the new set of values for $\sigma_{T_G}$ at 1073 K and 1119 K , recalculate $\sigma_{T_N}$ at 948 K , 998 K, 1011 K, and 1023 K using eq. (S-1), linearly extrapolate $\sigma_{T_N}$ to obtain another set of $\sigma_{T_G}$ . Continue this step until convergence is reached.

After convergence, the values of $\sigma_{T_N}$ at 948 K, 998 K, 1011 K, and 1023 K, and $\sigma_{T_G}$ at 1073 K and 1119 K were determined. The critical work of cluster formation, $W^*$, was calculated using eq. (S-2)[2], and the interfacial free energy, $\sigma$ , the Gibbs driving free energy per volume, $|\Delta g_{\mathrm{v}}|$ for each temperature.

$$W^* = \frac{16\pi}{3}\frac{\sigma^3}{|\Delta g_{\mathrm{v}}|^2} . \qquad \text{(S-2)}$$

The measured induction time, $\theta_{n^*(T_G)}$, was for the critical size at the growth temperature. To compare with the predictions of CNT, however, the induction time for the critical size at the nucleation treatment temperature, , $\theta_{n^*(T_N)}$ , is required. Following Xia *et al.*[1], $\theta_{n^*(T_N)}$ can be calculated from $\theta_{n^*(T_G)}$ using the following expression[2,3]

$$\frac{\theta_{n^*(T_G)}}{\theta_{n^*(T_N)}} = \frac{6}{\pi^2}\left[\xi + \ln\xi + \ln\left(\frac{6W^*}{k_\mathrm{B}T}\right) + \zeta_\mathrm{E} - 1\right], \tag{S-3}$$

where

$$\xi = \left(\frac{\sigma_{T_G}}{\sigma_{T_N}}\right)\Bigg/\left(\frac{|\Delta g_\mathrm{v}|_{T_G}}{|\Delta g_\mathrm{v}|_{T_N}}\right) - 1. \tag{S-4}$$

## References cited in the supplementary methods

1. Xia, X. *et al.* Time-dependent nucleation rate measurements in BaO·2SiO$_2$ and 5BaO·8SiO$_2$ glasses. *J. Non. Cryst. Solids* **525**, 119575 (2019).
2. Kelton, K. F. & Greer, A. L. *Nucleation in condensed matter: applications in materials and biology*, Elsevier, Amsterdam (2010).
3. Shneidman, V. A. & Weinberg, M. C. Induction time in transient nucleation theory. *J. Chem. Phys.* **97**, 3621–3628 (1992).